\documentclass[review, 10pt]{elsarticle}
\usepackage{booktabs,multirow,hhline}
\usepackage{amssymb}
\usepackage{geometry}
\usepackage{graphicx}
\usepackage{wrapfig}
\usepackage{pifont}
\usepackage{txfonts}
\usepackage{subfigure}
\usepackage{indentfirst}
\usepackage{lineno}
\usepackage{hyperref}
\usepackage{array}
\journal{Astroparticle Physics}
\newcommand{\head}[2]{\multicolumn{1}{|>{\centering\arraybackslash}p{#1}|}{\textbf{#2}}}
\usepackage[small]{caption} 
\newcommand{\dg}{$^\circ$}

\begin{document}
\begin{frontmatter}

\title{Magnetic Deflections of Ultra-High Energy Cosmic Rays\\ from Centaurus A}
\author{Azadeh Keivani$^{1,*}$, Glennys R. Farrar$^2$, and Michael Sutherland$^{1,3}$}
\address{$^1$Department of Physics \& Astronomy, Louisiana State University, Baton Rouge, LA\\ 
$^2$Center for Cosmology and Particle Physics, Department of Physics, New York University, New York, NY\\
$^3$Department of Physics \& Center for Cosmology and AstroParticle Physics, The Ohio State University, Columbus, OH\\
$^*$akeivani@phys.lsu.edu}

\begin{abstract}

We present the results of a study that simulates trajectories of ultra-high energy cosmic rays from Centaurus A to Earth, for 
particle rigidities from $E/Z = 2$ EV to 100 EV, i.e., covering the possibility of primary particles as heavy as Fe nuclei with energies exceeding 50 EeV. 
The Galactic magnetic field is modeled using the recent work of Jansson and Farrar (JF12) which fitted its parameters to match extragalactic Faraday rotation measures and WMAP7 synchrotron emission maps.  
We include the random component of the GMF using the JF12 3D model for $B_{\rm rand}(\vec{r})$ and explore the impact of different random realizations, coherence length and other features on cosmic ray deflections.  Gross aspects of the arrival direction distribution such as mean deflection and the RMS dispersion depend mainly on rigidity and differ relatively little from one realization to another.  However different realizations exhibit non-trivial substructure whose specific features vary considerably from one realization to another, especially for lower rigidities.  At the lowest rigidity of 2 EV, the distribution is broad enough that it might be compatible with a scenario in which Cen A is the principle source of all UHECRs.  No attempt is made here to formulate a robust test of this possibility, although some challenges to such a scenario are noted.
\end{abstract}

\begin{keyword}
Ultra-High Energy Cosmic Rays \sep Galactic Magnetic Fields \sep Centaurus A
\end{keyword}

\end{frontmatter}

\section{Introduction}

One of the biggest mysteries in particle astrophysics is the origin of ultra-high energy cosmic rays (UHECRs).
There are few source types that appear capable of accelerating the particles to energies as high as has been observed $(>10^{20}$ eV).
Moreover, UHECRs with energies above $\sim5 \times 10^{19}$ eV (50 EeV) lose energy through interactions with cosmic microwave background photons so that any that are observed
must have originated within about 200 Mpc \cite{zk,g}.
Centaurus A (Cen A) is the closest active galactic nucleus (AGN) to Earth and is thus one of the prime source candidates.
The Pierre Auger Observatory has reported a modest excess of cosmic ray arrival directions from within about 20$^\circ$ of the celestial location of Cen A in its UHECR data~\cite{Auger2010}.
This further increases the interest to study the propagation of UHECRs through the Galactic magnetic fields (GMF) from Cen A.

UHECRs are mostly or entirely charged particles \cite{Auger2009} so they are deflected in the magnetic fields that they traverse on their way to Earth.
In this study, we simulate trajectories of cosmic rays from Cen A to Earth to examine the deflections that occur due to the GMF and turbulent field realizations. 
We construct the GMF according to the recent model of Jansson and Farrar \cite{JF12model,JF12random} (JF12).
The functional form of the field used in JF12 model is more general than previously considered and is tuned to fit the observed rotation measure and polarized synchrotron data.
An overview of the model is given in Section \ref{JF12model}.

This study is an extension of the work of Farrar {\it et al.} \cite{FJFG} which simulated UHECRs from the center of Cen A considering only the coherent field and four rigidities (energy divided by the particle's charge: $R=E/Z$) of 160, 80, 40 and 20 EV.
Here we simulate additional rigidities extending down to 2 EV from the center of Cen A as well as consider the northern and southern radio lobes and explore the impact of adding several realizations of a Kolmogorov random field to the coherent JF12 GMF model.

Cen A was proposed to be the possible source of most cosmic rays by Farrar and Piran \cite{FP} assuming constraints on the values of turbulent extragalactic magnetic fields; see also \cite{Ginzburg}.
Since the Pierre Auger Observatory released its first set of UHECR events, there have been several studies on the correlation of UHECRs with Cen A.
For example, Gorbunov {\it et al.} discuss the possibility that Cen A could be the source of a significant portion of the observed events \cite{Gorbunov}.
Wibig and Wolfendale assume that all observed UHECRs originate from three sources including Cen A \cite{Wibig}.
Hardcastle {\it et al.} argue the possibility that the giant radio lobes of Cen A can accelerate UHECRs \cite{Hardcastle}.
Related works include \cite{Kachelriess,Rachen,Fargion1,Fargion2,Rieger,Moskalenko,Nagar}.
Ref. \cite{Giacinti} also examined the effects of turbulence on UHECRs, however with an older model of the GMF and turbulent realizations that were not constrained by data. 

The propagation method is briefly explained in Section \ref{tracking}, some properties of Cen A are given in Section \ref{cena}, and finally the results of the simulation are discussed in Sections \ref{regular} and \ref{krf}.

\section{JF12 GMF model}
\label{JF12model}
JF12 is a recent GMF model that was fit to the WMAP7 Galactic synchrotron emission maps ($\approx $120,000 pixels of data) \cite{wmap} and more than forty thousand extragalactic rotation measures to constrain its parameters.
This model includes a large scale regular component, a so-called ``striated" random component that is found to follow the larger geometric features of the regular component \cite{JF12model}, and a turbulent random field component \cite{JF12random}. The random component of the JF12 model is constrained by the WMAP total Galactic synchrotron emission map \cite{JF12random}.
 
In this study, we first consider the large scale regular component of the JF12 model.
The component contains a disk field component, a toroidal halo component and an out-of-plane component as suggested by observations of external galaxies.
Then, for a subset of rigidities, we add to this several realizations of a turbulent random field whose r.m.s value is that of the JF12 random field model \cite{JF12random}, each characterized by a Kolmogorov power spectrum with different maximum coherence scales. 
The striated field is not included.  Its effect is analogous to smearing the cosmic ray rigidity, i.e., it smears arrival directions along the arc predicted as a function of rigidity for the coherent field alone.  Using the striation parameter determined in \cite{JF12random},  the smearing from the striated field can be estimated to be $\simeq 14\% \left(\frac{\lambda/100\, {\rm pc})}{D/10 \,{\rm kpc}}\right)^{\frac{1}{2}}$ of the deflection in the coherent field, where $D$ is the distance propagated through the region of significant $B$ and $\lambda$ is the typical coherence length of the striated random field.

We ignore the extragalactic magnetic field in this study because Cen A is relatively close to us and the strength of the extragalactic magnetic field is expected to be much smaller than of the GMF \cite{dolag04,dar}.  
The typical cosmic ray deflection magnitude in a turbulent extragalactic magnetic field ($\delta\theta_{\rm EG}$) can be estimated \cite{FJFG} to be,
\begin{equation}
\label{EGMF}
\delta\theta_{\rm EG}\approx0.15^\circ\left({\frac{D}{3.8\ {\rm Mpc}}}\cdot{\frac{\lambda_{\rm{EG}}}{100\ \rm{kpc}}}\right)^{\frac{1}{2}} \left(\frac{B_{\rm EG}}{1\ \rm nG}\right) \left(\frac{Z}{E_{100}}\right),
\end{equation}
\noindent where $D$ is the source distance, $B_{\rm EG}$ and $\lambda_{\rm EG}$ are the r.m.s value and coherence length of the magnetic field, $Z$ is the charge of the UHECR in units of the proton charge, and $E_{100}$ is the UHECR energy in units of 100 EeV.
A 100 EeV proton originating from Cen A would thus be deflected by only $\approx0.15^\circ$ in such a turbulent extragalactic magnetic field, which is small compared to that from the regular and random Galactic fields of JF12 determined below.

\section{Tracking}
\label{tracking}
We use the latest release (v3.0.4) of \textit{CRT} \cite{sutherland} to inject and propagate UHECRs in the JF12 model. 
\textit{CRT} uses adaptive Runge-Kutta integration methods to determine the trajectory of a charged particle through a magnetic field according to the relativistic Lorentz force.

The method we use to track the cosmic rays through the GMF is ``backtracking" particles from the Earth and recording their locations and directions when they exit the Galactic field region.
This is accomplished by following the antiparticle, which simply means changing the sign of the intended particle's charge and reversing its incoming velocity vector.  The propagation of a particle stops when it hits the pre-defined boundary of the Galaxy since the extragalactic magnetic field is assumed to have negligible magnitude.

\begin{wrapfigure}{l}{0.42\textwidth}
\vspace{-0.15in}
\centering
\includegraphics[width=0.4\textwidth]{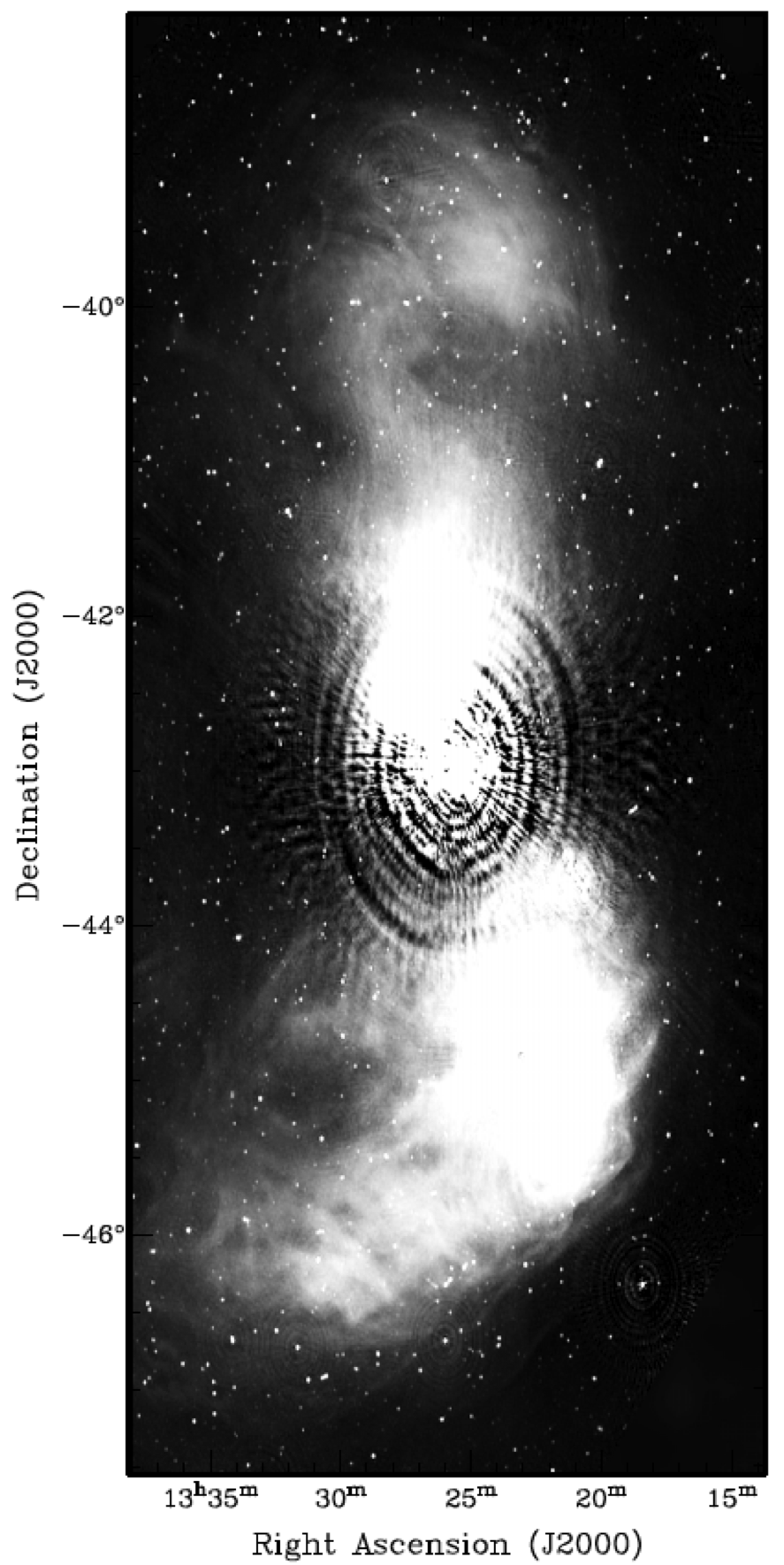}
\vspace{-0.2in}
\caption{High resolution radio image of Cen A at 1.4 GHz showing the center of Cen A and its radio lobes in equatorial coordinates. The northern lobe is at the top of the image and the southern lobe is at the bottom. Image is Figure 3 of \cite{Feain-etal}.}
\label{CenAimage}
\vspace{-0.05in}
\end{wrapfigure}

We backtrack cosmic rays from isotropically distributed initial locations using HEALPix \cite{healpix}.
In this study, we pick HEALPix resolution index 11 (or 9) which tiles the sky with more than $5\times 10^{7}$ (or $3\times10^6$) pixels, each covering $8 \times 10^{-4}$ deg$^{2}$ (or $1.3 \times 10^{-2}$ deg$^{2}$).  
We backtrack one particle with a fixed rigidity $R$ from the center of each pixel.  
The Larmor radius of a relativistic charged particle in a magnetic field $B$ depends only on its rigidity, so the use of rigidity allows a single simulation to represent a variety of primary particle compositions with appropriately scaled total primary energies.
An event with rigidity $R$ could be a proton with energy $R$ or an iron nucleus with energy $26R$.
Rigidities considered start with log($R_{\rm{EV}}$)=2 (or $R=100$ EV) and end at log($R_{\rm{EV}}$)=0.3 (or $R \simeq 2.0$ EV); for the coherent field, steps of 0.02-0.05 in log($R_{\rm{EV}}$) are used.
When the random field is included the simulations are significantly more time-consuming therefore only four values of rigidity have been used:  log($R_{\rm{EV}}$)=2, 1.5, 1.0, and 0.5 (or $R=100, 31.62, 10, 3.16$ EV), respectively.
For each rigidity, we identify particles that backtrack to regions of a specified solid angle within a radius of 1$^{\circ}$ or 3$^{\circ}$ around a chosen source location; in the first case this is the center of Cen A.  
We repeat the process for source locations centered on the tips of the northern and southern radio lobe.

\section{Centaurus A}
\label{cena}
Centaurus A (NGC 5128) is a Fanaroff-Riley class I \cite{FanaroffRiley} massive elliptical radio galaxy in the constellation of Centaurus.
It is located at a distance of $3.8\pm0.1$ Mpc \cite{Harris} and is the closest radio galaxy to Earth. It has a radio luminosity of $ L_{1.4\ \rm{GHz}}=2.3\times 10^{24}\ \rm{W\ Hz}^{-1}$ \cite{Cooper} and is
a uniquely well-studied source that has been characterized over the entire electromagnetic spectrum.
Cen A has an active galactic nucleus in its center which ejects matter in relativistic jets: a pair of asymmetric nuclear jets and a pair of inner jets \cite{Feain-etal}.
The inner jets extend about 1.4 kpc from the center \cite{Burns} whereas the inner radio lobes extend a further 5 kpc from the nucleus \cite{Clarke}.
Cen A has also middle and outer radio lobes that extend to about 14 and 500 kpc, respectively.

Figure~\ref{CenAimage} shows a high resolution radio image of Cen A~\cite{Feain-etal}.
The image has angular dimensions of $8^{\circ}\times 4^{\circ}$ translating to a physical size of 500 kpc $\times$ 200 kpc ~\cite{Israel} at the distance of Cen A.
The center of Cen A is taken to be $(l,b)=(-50.5^\circ, 19.4^\circ)$ in Galactic coordinates.
We define $(l,b)=(-49.8^\circ, 24.4^\circ)$ and $(-51.3^\circ, 14.5^\circ)$ to be the tip of the northern and southern radio lobes, respectively.
In equatorial coordinates these directions correspond to $(\rm 13^h 25^m 3.9^s, -38.0^\circ)$, $(\rm 13^h 25^m 23.5^s, -43.0^\circ)$, and $(\rm 13^h 24^m 52.9^s, -48.0^\circ)$.

\section{UHECR Deflections in JF12 Coherent Field}
\label{regular}
\begin{figure}[!ht]
\begin{center}
\subfigure[$R=$ 2 - 100 EV]{
\includegraphics [trim=1.4cm 0cm 0cm 0cm,scale=0.38]{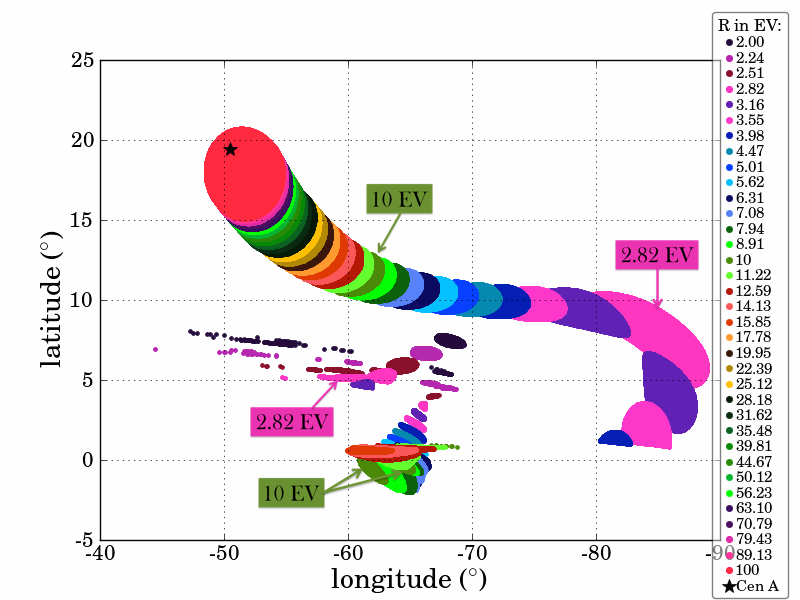}}
\subfigure[$R = 56.23$ EV]{
\includegraphics [trim=0.7cm 0cm 1cm 0cm,scale=0.38]{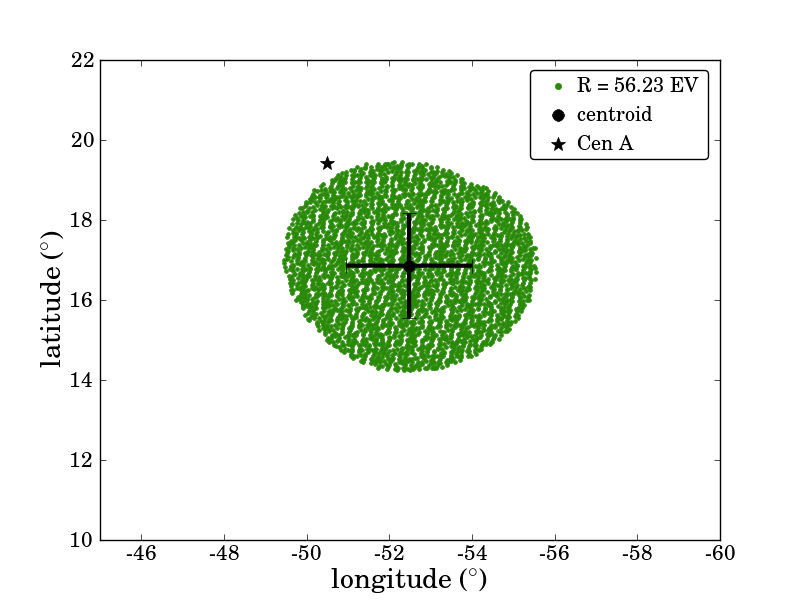}
\label{lbexample}}
\caption{Arrival directions of the simulated events originating from within 3$^{\circ}$ around the center of Cen A, in the JF12 coherent field, for (a) $R=2$ EV to 100 EV and (b) focussing on $R = 56.23$ EV, for which the centroid and standard deviations of the direction distribution are shown by the dot and error bars.}
\label{snakeplot}
\end{center}
\end{figure}

\begin{figure}[ht!]
\begin{center}
\includegraphics [trim=0cm 0cm 0cm 0cm,width=1\textwidth]{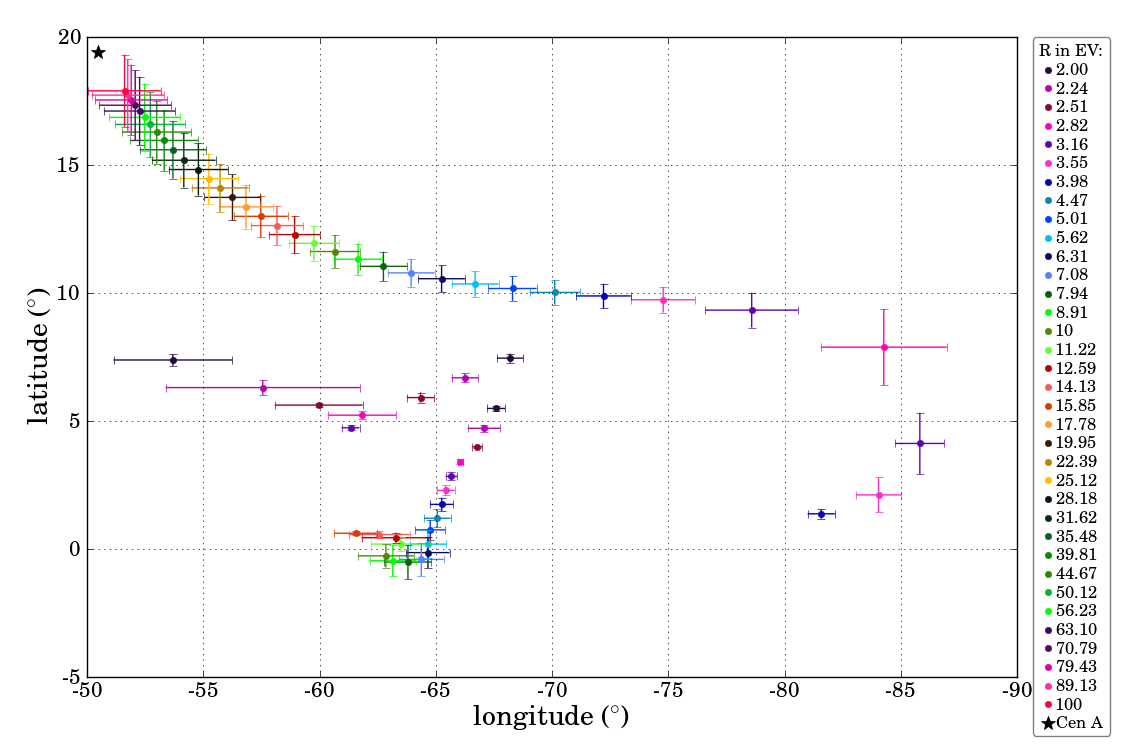}
\caption{Centroids and standard deviations of the arrival direction distributions of events with the same rigidity. Their source directions lie within 3$^{\circ}$ of the center of Cen A. The map is projected in Galactic coordinates.}
\label{fwdbkcentroid3deg}
\end{center}
\end{figure}

In the following we consider UHECRs originating within $3^{\circ}$ or within $1^{\circ}$ of the center of Cen A, for different cases.
For the lower resolution HEALPix level 9 simulations we use the $3^{\circ}$ locus.
The HEALPix level 11 simulations give us a similar event statistics with a $1^{\circ}$ radius.
A $3^{\circ}$ angular window encompasses a sizeable portion of the radio galaxy (see Figure~\ref{CenAimage}) which may be sites of acceleration and emission of cosmic rays.
Since Cen A is about 3.8 Mpc away from the Earth, the $3^{\circ}$ ($1^{\circ}$) radius circle encompasses a physical distance of about 200 (66) kpc in radius from the center.
As discussed earlier, the limits of the outer radio lobes lie about 500 kpc from the center so a value of 200 (66) kpc is a reasonable estimate for including acceleration regions associated with the central engine.

Figure~\ref{snakeplot}(a) shows the arrival direction distribution of simulated events from within $3^{\circ}$ of the center of Cen A, for different rigidities.
We calculate the longitude and latitude centroids of the arrival direction distributions as well as standard deviations in each direction separately; these are shown as black points and lines respectively for $R = 56.23$ EV in Figure~\ref{lbexample}.
The defined source region is a circular $3^{\circ}$ patch on the sky and the arrival distribution generally retains this shape.

The process is repeated for 35 different values of rigidity for the JF12 coherent field.
Figure~\ref{fwdbkcentroid3deg} shows the centroids and standard deviations of the arrival direction distributions from the center of Cen A for all simulated rigidities.
Each color represents a different rigidity as indicated in the legend.
These points illustrate how the arrival directions of UHECRs from the center of Cen A deviate from the source direction as a function of the rigidity of the particle.
Particles with large $R$ values deviate much less than those with smaller values.
Note that for $R < 17.78$ EV, the arrival directions branch into two distinct regions on the plot; the centroid and standard deviations of such split distributions are computed and plotted separately.
The distributions resemble the circular nature of the defined source region down to low simulated rigidities ($R < 3$ EV) below which the distributions deform into clearly noncircular regions.
None of the arrival distributions has a significant number of outliers that would skew the appearance of Figure~\ref{fwdbkcentroid3deg}; the bars accurately reflect the distributions.
Note that the regions for $R=3.98$ EV and 3.55 EV at the bottom-right side of Figure~\ref{snakeplot}(a) appear to be cut but this is a real effect and the events with slightly smaller latitudes backtrack to a region to the northwest of Cen A and lie outside of the $3^{\circ}$ radius circle. 

We separately study the observed directions of the simulated events coming from within 3$^\circ$ near the tips of the northern and southern radio lobes.
We find that the general stream of the events are the same, but the appearance of separated regions split away from the main thread begins at lower rigidities ($R=12.59$ EV) for the northern tip  source and at higher rigidities ($R=22.39$ EV) for the southern tip source.
The main thread is shifted towards greater latitude values for the source at the northern tip and is shifted towards smaller latitude values for the source at the southern tip compared to the source at the center of Cen A.
These shifts are only a few degrees for the lower rigidities.
The shape and orientation of the threads remain the same as for the Cen A center source case in the sense that progressively lower rigidities trail towards lower longitude and latitude values.
The multiple images observed for lower rigidities shift primarily in longitude by about ten degrees but maintain the same curved rigidity ordering and structure.

\section{UHECR Deflections in JF12 GMF Model including Random Fields}
\label{krf}

\begin{figure}[!hb]
\centering
\subfigure[$R=100$ EV]{
\includegraphics[width=0.45\textwidth,trim=1cm 0cm 1cm 0cm]{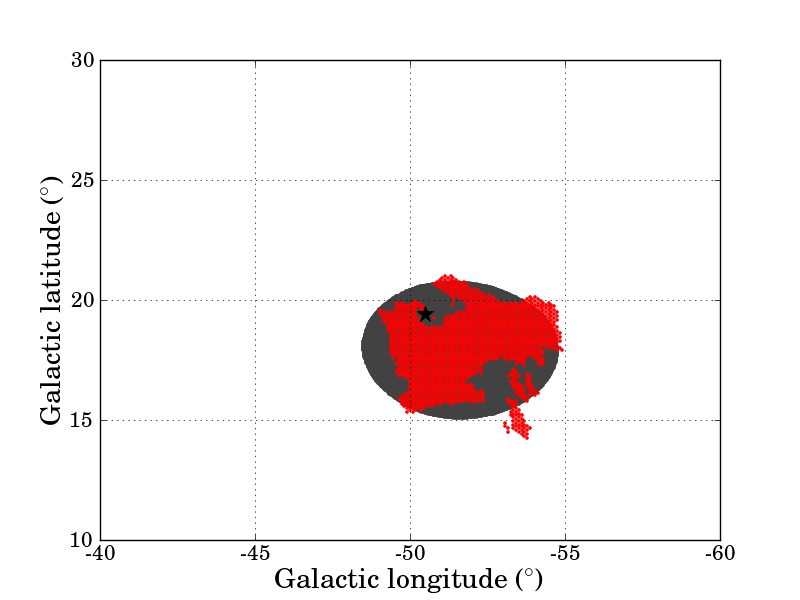}
\label{krf1-r100-3deg}}
\subfigure[$R=31.62$ EV]{
\includegraphics[width=0.45\textwidth,trim=1cm 0cm 1cm 0cm]{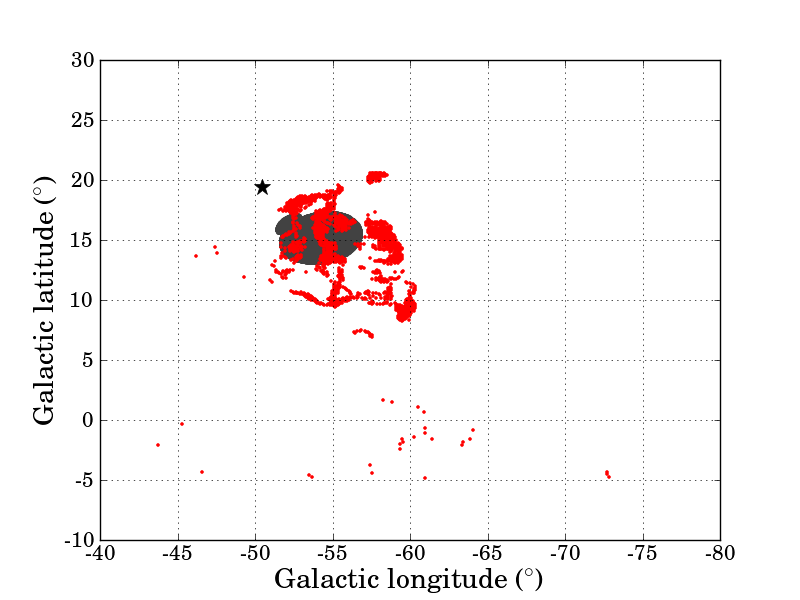}}
\subfigure[$R=10$ EV]{
\includegraphics[width=0.45\textwidth,trim=1cm 0cm 1cm 0cm]{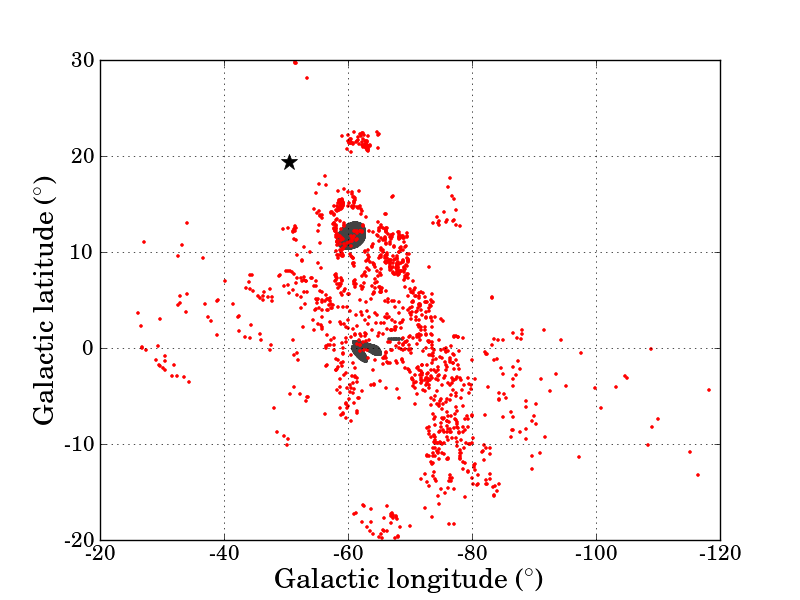}}
\subfigure[$R=3.16$ EV]{
\includegraphics[width=0.45\textwidth,trim=1cm 0cm 1cm 0cm]{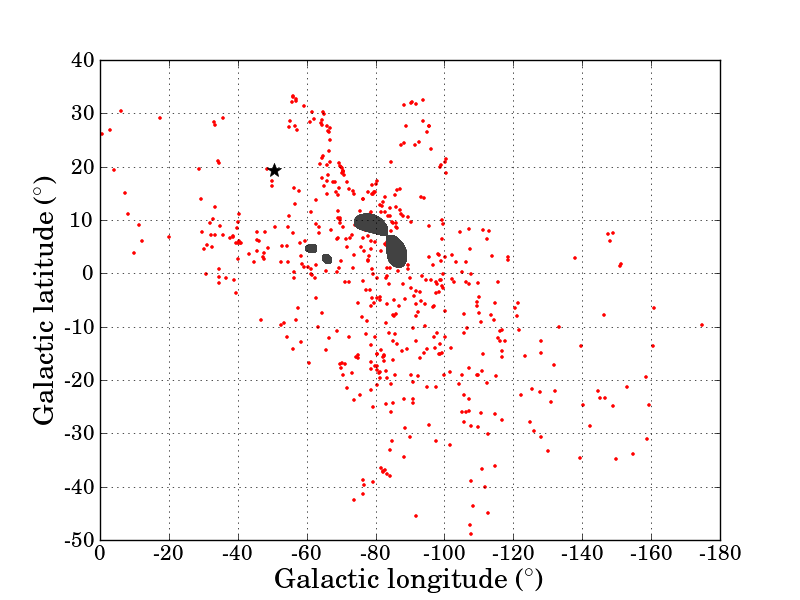}}
\caption{Rigidity dependence of deflections of cosmic rays within $3^{\circ}$ of the center of Cen A with regular (grey) and regular plus KRF1 realization (red).  
The black asterisk shows the location of the center of Cen A.}
\label{krf1-rall-3deg}
\end{figure}

In this section we discuss the effects of adding the detailed Kolmogorov random field component to the JF12 regular component.   UHECR propagation in the full GMF including the random component requires an explicit realization of the field, so we have developed a method for propagation by implementing a box containing a grid of field magnitudes that follow a Kolmogorov spectrum \cite{AK-thesis}.
This box is repeated throughout the GMF volume and is locally scaled by the JF12 r.m.s random field strength at the point of interest, i.e., the particle position. 
The total magnetic field is the vector sum of the regular and random field vectors.
Independent random field realizations can be constructed by generating boxes with different random number seeds and different maximum coherence lengths.

Since the simulations are computationally expensive we first consider the regular field plus one random field realization (called KRF1), backtracking one event per HEALPix resolution index 9 pixel ($3 \, 10^6$ events over the sky), for four different rigidities: $R = 100, 31.62, 10,$ and 3.16 EV.
The arrival direction distributions of events originating within $3^\circ$ of the center of Cen A in the regular field only and regular plus KRF1 field are compared in Figure \ref{krf1-rall-3deg}.
We see that the results of propagation only begin to change significantly at lower rigidity, such as 31.62 EV, between the regular-only and the regular-plus-random models.  Even at lower rigidities, the arrival distribution centroids with and without the random field are only separated by a few degrees, but a large broadening of the distribution is evident when the random field is included.

Next, we consider a suite of random field realizations, to examine the effects of random field characteristics on the arrival direction distributions.  We do this for a single rigidity which we choose to be $R=10$ EV.  At $R=10$ EV the effects of the random field are clearly important while the dispersion remains small enough that the study is computationally feasible. For these simulations we increase the HEALPix resolution index to 11 ($5 \, 10^7$ events all-sky); this enables us to reduce the angular size of the source region to 1$^\circ$ instead of 3$^\circ$ around the center of Cen A and maintain an adequate number of events in the sample.
\begin{table}[!b]
\vspace{-0.05in}
\footnotesize
\centering
\caption{Properties of each random field realization used in this paper. 
Column 2 shows the $\lambda_{\rm{max}}$ used in each field realization; 
Col. 3 indicates whether the realization box is fixed or randomly flipped throughout the Galactic volume;
Cols. 4 and 5 are the HEALPix resolution index and the angular radius around the center of Cen A from which we select event source directions.
Col. 6 shows the rigidity for the given simulation. 
Col. 7 gives the total number of received events from the given source locus; in the absence of magnetic lensing this would be 2155 and 3832 for res 9 in 3\dg\ and res 11 in 1\dg\, respectively.
Col. 8 gives the mean deflection $\Delta$ with respect to the source direction, and the standard deviation of the angular separation of the events relative to that mean.
Columns 9 and 10 give the centroid and width of the arrival direction distribution in Galactic coordinates.}
\begin{tabular}{|c|cccc|c|c|c|cc|}
\hline
\head{0.8cm}{Realization} & \head{0.7cm}{$\lambda_{\rm{max}}$ (pc)} & \head{1.1cm}{Orient. Method} & \head{0.5cm}{Res.} & \head{0.5cm}{$\Psi$ (deg)}  & \head{0.7cm}{$R$ (EV)} & \head{0.7cm}{$\#$ of Evts} & \head{0.7cm}{$\Delta\ \rm{(SD)}$ (deg)} & \head{1.4cm}{$\bar{l}\pm$STDV (deg)} & \head{1.4cm}{$\bar{b}\pm$STDV (deg)}\\
\hline
\multicolumn{0}{|c|}{\multirow{4}{*}{KRF1}} &
\multicolumn{0}{c}{\multirow{4}{*}{100}} &
\multicolumn{0}{c}{\multirow{4}{*}{flipped}} &
\multicolumn{0}{c}{\multirow{4}{*}{9}} &
\multicolumn{0}{c|}{\multirow{4}{*}{$3^\circ$}} & 3.16 & 540 & 44.6 (19.8) & $-79.5\pm37.9$ & $-1.5\pm17.6$ \\
\cline{6-10}         &     \multicolumn{4}{c|}{} &10 & 1349 & 24.7 (8.8) & $-66.5\pm12.6$ & $2.9\pm9.5$ \\ 
\cline{6-10}         &     \multicolumn{4}{c|}{} & 31.6 & 1543 & 8.1 (2.4) & $-56.1\pm2.6$ & $13.9\pm3.8$ \\ 
\cline{6-10}         &     \multicolumn{4}{c|}{} &100 & 1421 & 2.5 (0.8) & $-51.9\pm1.5$ & $18.0\pm1.4$  \\ 
\hline
\hline
KRF1 & 100 & flipped & 11 & $1^\circ$ & 10 & 2292 & 24.7 (7.9) & $-66.7\pm11.7$ & $2.7\pm9.5$ \\
\hline
KRF2 & 100 & flipped & 11 & $1^\circ$ & 10 & 1877 & 20.7 (7.2) & $-63.5\pm10.7$ & $5.9\pm8.0$ \\
\hline
KRF2 & 100 & fixed & 11 & $1^\circ$ & 10 & 1827 & 21.6 (6.4) & $-66.1\pm10.9$ & $5.8\pm8.2$ \\
\hline
\multicolumn{0}{|c|}{\multirow{5}{*}{KRF3}} &
\multicolumn{0}{c}{\multirow{5}{*}{512}} &
\multicolumn{0}{c}{\multirow{5}{*}{fixed}} &
\multicolumn{0}{c}{\multirow{5}{*}{11}} &
\multicolumn{0}{c|}{\multirow{5}{*}{$1^\circ$}} & 2 & 679 & 58.4 (32.5) & $-40.8\pm68.1$ & $-8.2\pm29.7$ \\
\cline{6-10} 	  &	 \multicolumn{4}{c|}{} & 3.16 & 723 & 42.2 (24.6) & $-54.7\pm44.4$ & $-3.3\pm23.2$ \\
\cline{6-10}         &     \multicolumn{4}{c|}{} &10 &  800 & 27.7 (13.9) & $-62.4\pm17.8$ & $-1.0\pm6.2$ \\ 
\cline{6-10}         &     \multicolumn{4}{c|}{} & 31.6 & 1592 & 14.8 (3.2) & $-57.4\pm3.5$ & $6.6\pm6.3$ \\ 
\cline{6-10}         &     \multicolumn{4}{c|}{} & 100 & 4719 & 1.8 (0.4) & $-50.6\pm0.8$ & $17.8\pm0.6$ \\ 
\hline
KRF4 & 512 & fixed & 11 & $1^\circ$ & 10 & 1037 & 33.4 (12.4) & $-70.7\pm20.6$ & $-1.6\pm9.4$ \\
\hline
GARF & 512 & fixed & 11 & $1^\circ$ & 10 & 2067 & 19 (9.1) & $-60.2\pm14.6$ & $7.5\pm9.4$ \\
\hline
\hline
KRF3-disk/3 & 512 & fixed & 11 & $1^\circ$ & 10 & 1329 & 25.3 (5.0) & $-66.6\pm7.7$ & $0.9\pm$4.7\\
\hline
\end{tabular}
\label{realizations}
\end{table}

\begin{figure}[ht]
\centering
\includegraphics[width=0.5\textwidth,trim=1cm 0cm 1cm 0cm]{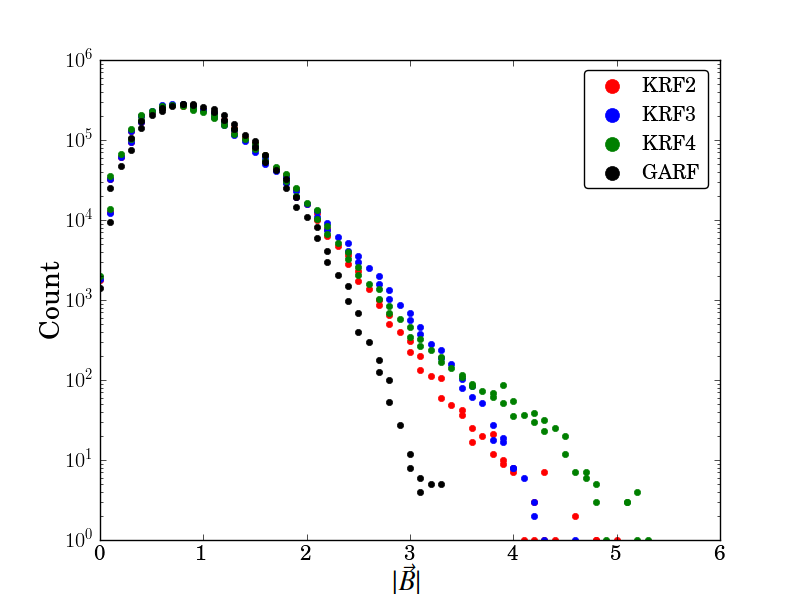}
\caption{Histogram of field strength in the unit box for KRF2, KRF3, KRF4 and GARF.}
\label{Btot}
\end{figure}

Four realizations of the JF12 random field are generated using \textit{CRT} code (KRF1, KRF2, KRF3, KRF4).
Two of these realizations (KRF2 and KRF3) are implemented multiple times as will be discussed below.
A single realization (GARF) is produced using the commonly used GARFIELD method \cite{garfield} in order to compare different field generation implementations.  A final realization (KRF-disk/3) reduces by a factor of 3 the strength of the disk component of the JF12 random field, motivated by the estimates in \cite{GF-CRAS} of the potential impact of replacing the GALPROP cosmic ray electron model used in JF12 by a more structured $n_{\rm cre}$ such as predicted in some recent source models such as \cite{benyamin+13}.   A summary of the realizations and simulations is given in Table~\ref{realizations}.  The fields and results are explained in greater detail below.\\

\noindent {\bf{Comparing Realizations}}

We can characterize the different random field realizations by histograms of the magnetic field strength at each grid point in the unit box, before scaling with the JF12 r.m.s field strengths.
The field strengths of the unit box are normalized so that their r.m.s value is one. 
Figure~\ref{Btot} shows the field strength histogram for KRF2, KRF4 and GARF. 
It should be noted that KRF3 and KRF4 have the same field parameters, KRF2 has a lower $\lambda_{\rm{max}}$ than the other realizations, and GARF has the same parameters as KRF3 and KRF4 but is generated by a different method. 

Table \ref{realizations} lists specific details of and differences between the realizations, as described in the caption.
The number of events arriving at Earth that originate within the source locus is an interesting diagnostic.   In the absence of magnetic lensing, this number would simply be the number of events originating within the source locus: 2155 for 3\dg\ and res 9 and 3832 for 1\dg\ and res 11.  However due to magnetic lensing, the number varies with the field model and rigidity, as recorded in Column 7 of Table \ref{realizations}.  The mean angular deflections and RMS angular size of the arrival direction distribution, and the mean and standard deviations of the longitude and latitude of the events, are also listed in Table~\ref{realizations} for each simulation.

Taken together, the suite of realizations allow us to investigate how the following impact the arrival direction distributions: A) re-orientation or not of a realization within the simulation volume, B) variations between independent random realizations of fields with the same parameters,  C) maximum coherence length, and D) method of producing the Kolmogorov spectrum:

\begin{figure}
\centering
\includegraphics[width=0.45\textwidth,trim=1cm 0cm 1cm 0cm] {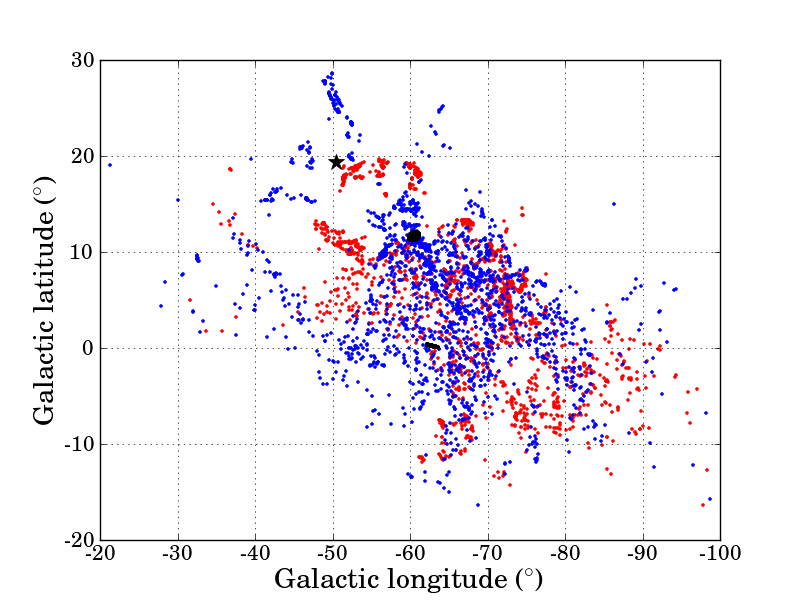}
\quad
\includegraphics[width=0.45 \textwidth,trim=1cm 0cm 1cm 0cm]{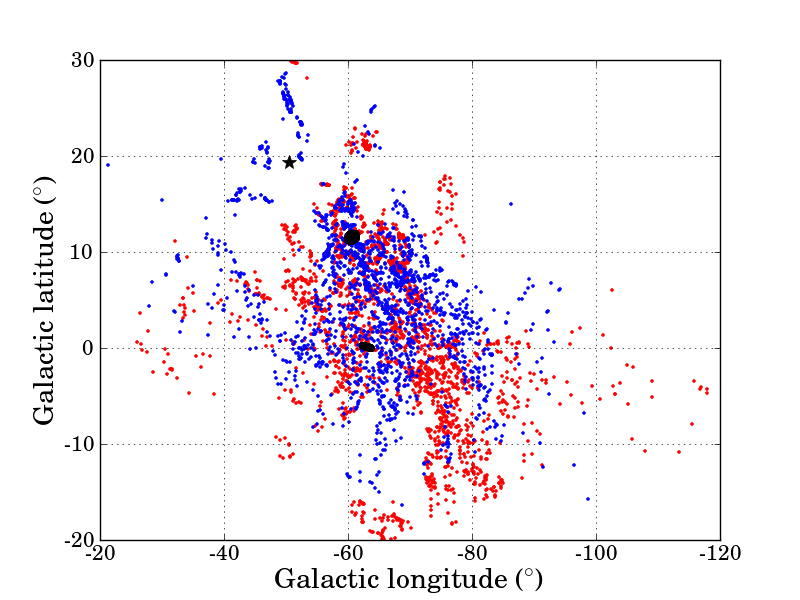}
\caption{Arrival direction distribution of events with $R=10$ EV, for source locations lying within $1^{\circ}$ of the center of Cen A;  the black regions give the direction distribution in the regular only field and the black star shows the location of the center of Cen A.  Left: KRF2 used in fixed orientation throughout the entire volume versus re-orienting different instances when filling the volume; KRF2-flipped in blue and KRF2-fixed in red.   Right: The effect of different random number seeds with the same procedure in other respects;  KRF1 in red and KRF2 in blue. } \label{krf1-krf2}
\vspace{-0.05in}
\end{figure}

\noindent{\bf{Independent Realizations with Same Parameters}}

KRF2 is used in a pair of simulations where the orientation of the realization box is either held fixed throughout the entire volume or is randomly re-oriented (flipped) from one location to another.
The left panel of Fig. \ref{krf1-krf2} illustrates the effect of different orientation methods implemented for a single realization.  The gross features of size and location of the arrival direction distribution is relatively insensitive to the orientation method (c.f.,Table~\ref{realizations}).

\begin{figure}
\centering
\includegraphics[width=0.45\textwidth,trim=1cm 0cm 1cm 0cm] {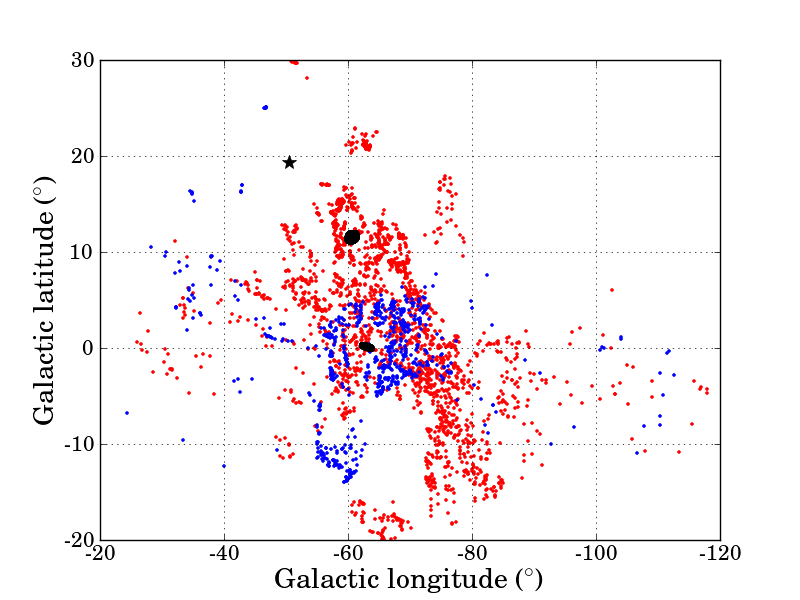}
\quad
\includegraphics[width=0.45\textwidth,trim=1cm 0cm 1cm 0cm] {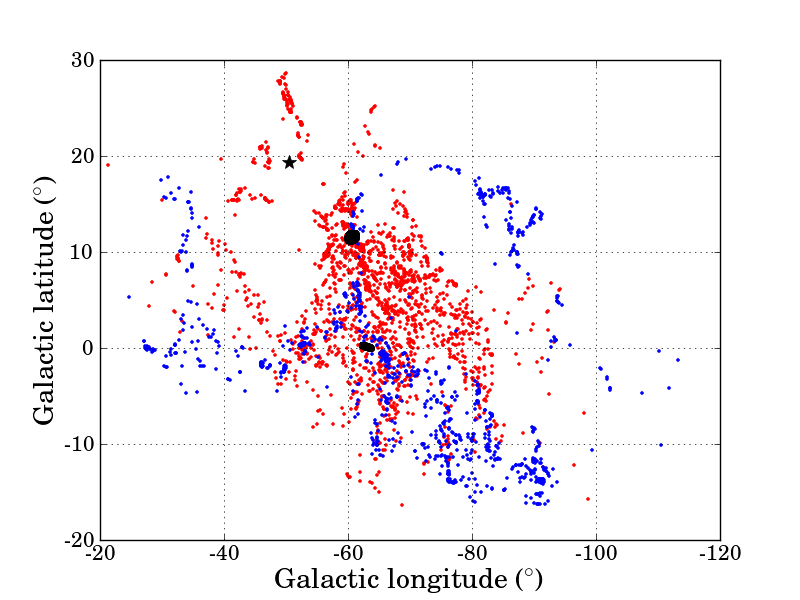}
\caption{Sensitivity to coherence length with $\lambda_{\rm max} = 100$ pc in red and $\lambda_{\rm max} = 512$ pc in blue.  The left panel shows KRF1\&3 and the right panel KRF2\&4; otherwise as in Fig. \ref{krf1-krf2}.  } \label{lambda}
\vspace{-0.05in}
\end{figure}

The right panel of Fig. \ref{krf1-krf2} compares the arrival directions for KRF1 and KRF2 to see the effect of two independent realizations (random number seeds) with the same random field parameters.
The last two columns of Table~\ref{realizations} show that these two distributions are only different up to statistical fluctuations.

\noindent{\bf{Sensitivity to Maximum Coherence Length and Random Field Generation}}

Maximum coherence length is related to the maximum wavelength of the Fourier modes of the random field realization, $\lambda_{\rm{max}}$.  According to Equation 2.6 of \cite{HMRS2002}, for a Kolmogorov spectrum and ample range between $k_{\rm min}$ and $k_{\rm max}$,  the coherence length is $\approx \frac{1}{5} \,\lambda_{\rm{max}}$ so that our choices of $\lambda_{\rm max}$ allow us to span the range of $\approx$20-100 pc coherence length.  KRF1 and 2 are generated using $\lambda_{\rm{max}}=100$ pc and KRF3 and 4 with 512 pc.
Figure \ref{lambda} show comparisons between KRF1 and KRF3, and KRF2 and KRF4, respectively, illustrating the effect of varying the maximum coherence length on the arrival direction distribution for 10 EV rigidity.  As expected, the differences from one realization to another are smaller for the smaller coherence length.  

\begin{figure}
\centering
\includegraphics[width=0.45\textwidth,trim=1cm 0cm 1cm 0cm] {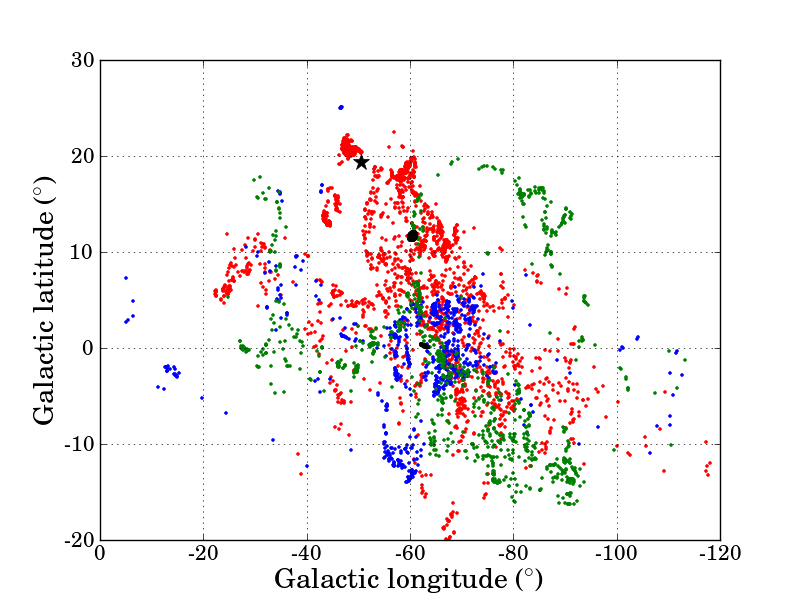}
\quad
\includegraphics[width=0.45\textwidth,trim=1cm 0cm 1cm 0cm] {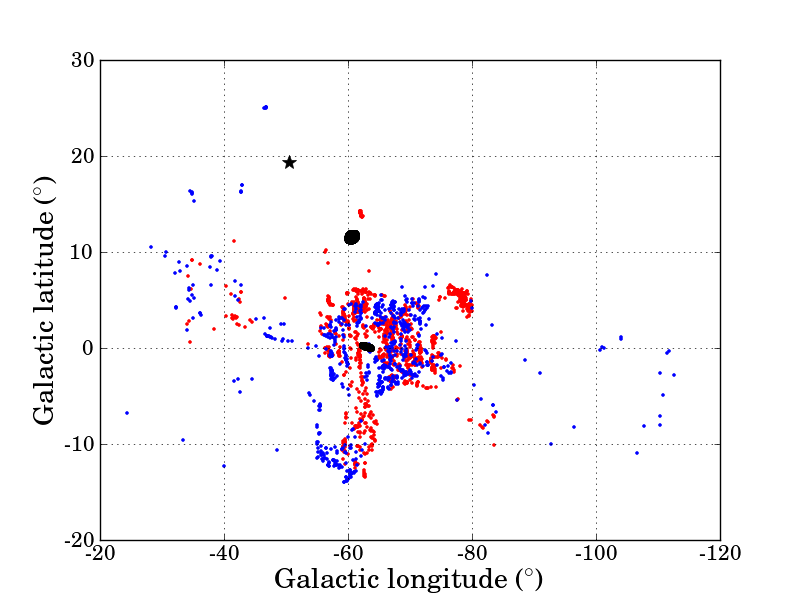}
\caption{Left panel: sensitivity to random field production method seen comparing use of GARF (red), KRF3 (blue) and KRF4 (green). Right panel: sensitivity to the RMS strength of the disk component of the random field; blue points are standard KRF3 and red points are with 3-times smaller random field in the disk (KRF3-disk/3). } \label{GARF&reduced}
\vspace{-0.05in}
\end{figure}

We can study the sensitivity to the tails of the distribution of $|B_{\rm rand}|$ values, by comparing KRF3 and KRF4 to GARF, since the coherence length (512 pc) is the same for each of these; the box orientation is also fixed throughout the simulation volume for all three realizations.  The left panel of Fig. \ref{GARF&reduced} shows arrival direction distributions for KRF3, KRF4 and GARF.
We see that GARF produces less dispersion, which is not surprising given its narrower tails in  $|B_{\rm rand}|$ values (Fig. \ref{Btot}). 

A larger coherence length, or broader $|B_{\rm rand}|$ distribution, seems to increase the ``cosmic variance'' from one realization to another.  It appears that for 10 EV UHECRs coming from the source direction of Cen A, differences from one realization to another can be sufficient to significantly impact the fraction of events whose trajectories come close to the Galactic plane, where the random field can be quite large in some arms \cite{JF12random}.  One sees from Table~\ref{realizations} that the variance about the mean deflection is much larger for the realizations in which the mean arrival direction is below the plane than for those above.  A subject of future work will be to see how this behavior changes for different source directions.
(Note that KRF1 was implemented with a flipped orientation scheme whereas the others use a fixed scheme, but this was found to make only a minor difference in the distribution.)

\noindent{\bf{Reduced random field in the Disk}}

Finally we see that reducing the disk random field strength by a factor of 3 relative to the JF12 random field model has little impact on the mean deflection but significantly decreases the arrival-direction spread.  This study was motivated to examine sensitivity of Cen A UHECR deflections to uncertainties in the GMF model \cite{GF-CRAS}.

\section{Summary and Conclusions}

We have performed a detailed exploration of how the UHECR image of Cen A moves and fragments as a function of CR rigidity from 100 EV to 2 EV in the JF12 model of the GMF.  This study shows that in the coherent field alone, even for deflections which can be large, the size and shape of the image changes only mildly until rigidities are below a few EV, as seen in Figs.~\ref{snakeplot}(a) and ~\ref{fwdbkcentroid3deg}.  However the picture changes dramatically when random fields are included.  
To explore this, we have performed high angular resolution simulations for a series of fixed realizations of the random field, for a variety of random-field realizations, to learn how much the arrival direction distribution changes from one realization to another.
(In many earlier studies of magnetic deflections, the random field was generated ``on-the-fly'' during propagation, such that each CR experiences a different field, leading to unrealistic additional smearing and smoothing of the average deflection map.) These simulations lead us to several conclusions, applicable for rigidities such that the UHECRs come close enough to the Galactic plane to feel the relatively strong random field in the disk ($\approx 10$ EV for Cen A as the source).
We used the JF12 model which is surely imperfect, but can be a guide to general features which can be expected:
\begin{itemize}
\item At rigidities of $  \simeq 10$ EV and below, individual deflection magnitudes can be very large and the arrival direction distributions span a large fraction of the sky.
\item The mean deflection and the RMS spread in arrival directions for a given rigidity varies a few degrees from one realization to another.  These properties depend much more on the rigidity of the CR than on the specific realization of the random field.
\item The distribution of arrival directions has considerable structure: it has hot-spots and edges, and would not typically be well-described by a 2-D gaussian.
\item The location and features of the structures vary significantly from one realization to another, even though the gross properties (mean, RMS) are similar.  
\item The ``transmission efficiency'' of UHECRs from a given source -- the number of CRs from that source which reach Earth, relative to the number which would reach us in the absence of the GMF -- varies from one realization to another as well as with rigidity.
\end{itemize} 
\begin{figure}[ht!]
\centering
\includegraphics[width=1\textwidth,trim=2cm 1cm 0cm 0cm] {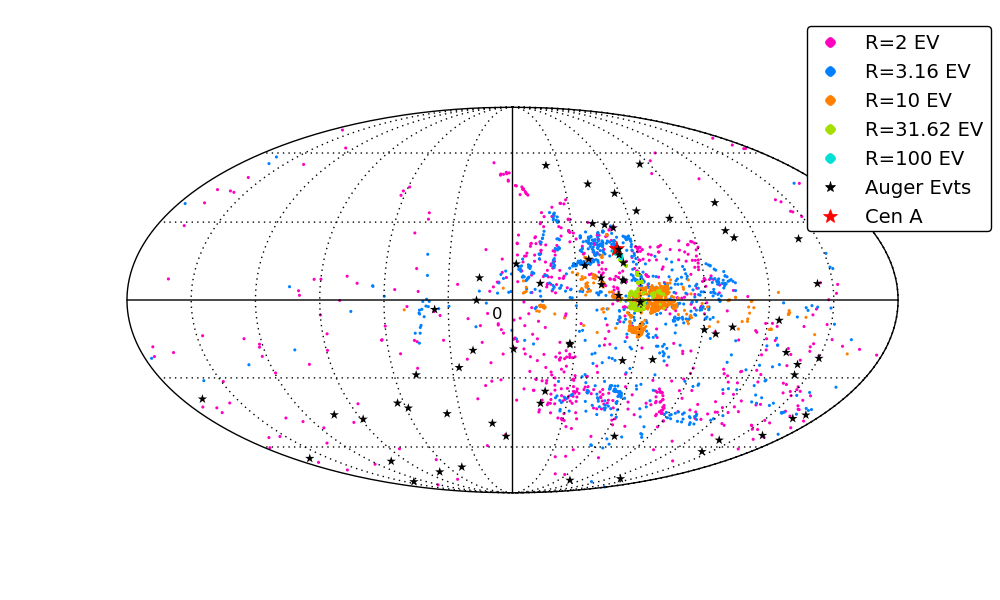}
\vspace{-0.5in}
\caption{Sky map in Galactic coordinates of the arrival direction distributions, for UHECRs from within $1^{\circ}$ of the center of Cen A, propagated through the regular JF12 GMF plus KRF3 realization of the random JF12 model, for rigidities from 2-100 EV.  
The observed Pierre Auger events with energies above 55 EeV (from~\cite{Auger2010}) are shown with black asterisks; note that the arrival locus for 100 EV is somewhat obscured by 2 Auger event symbols.}
\label{skymap-krf3-auger}
\end{figure}

We close by displaying (Fig. \ref{skymap-krf3-auger}) the simulated arrival direction distributions for a particular random field realization, for rigidities $R=2$, 3.2, 10, 32 and 100 EV, in a sky map using Mollweide projection.  
Observed events with $E \ge 55$ EeV published by the Pierre Auger Observatory are shown for comparison.  
The data are from~\cite{Auger2010}.
A striking but difficult to interpret feature of Fig. \ref{skymap-krf3-auger}, is that the simulated arrival direction distribution corresponding to $R =2$ EV, is so broad as to encompass most of the Auger events.  A detailed analysis taking into account the Auger exposure, including the recently-released TA events \cite{TAaniso14} and studying more realizations of the field model would be required to determine whether a single-source (Cen A) scenario could conceivably be viable.  Since photo-disintegration during propagation is unimportant because Cen A is so close, the UHECR spectral cut-off would reflect the maximum-rigidity that Cen A can produce.  At lower energies, one would expect a mixed composition reflecting the relative abundances at the source, with each component having a maximum energy $\approx (Z/26) \, E_{\rm cut-off} $.  A potential Achilles-heel of such a model is that --  no matter what the composition mix is, and even with a pure-Fe composition -- the events above 55 EeV must cover a range of rigidities extending at a minimum between $\approx 2$ and $\approx 6$ EV.  As can be seen from the distributions for $R=3.2$ and $R=10$ EV in Fig. \ref{skymap-krf3-auger}, events with these higher rigidities would be expected to be much more clustered than the $R \approx 2$ EV events, which may be incompatible with the observed approximate isotropy.   

In concluding, we stress that the JF12 random field model implemented here was only a first attempt to estimate the random component of the GMF.  As such, it is subject to significant revision as improved modeling and constraining data becomes available.  In particular, the possible very broad distribution of arrival directions of iron nuclei from Cen A our studies reveal, may not apply to the true GMF.   This underlines the importance of the observational challenge:  to find a robust way to test such scenarios, given that the specifics of the structure due to the random magnetic fields cannot be predicted in detail.  Information from individual event charge-assignments and from extending the energy range at which the  $X_{\rm max}$ and $X_{\mu,\rm max}$ distributions give composition information, for example as envisaged in the Auger Upgrade proposal, will give valuable complementary constraints to arrival-direction studies. \\

{\noindent\bf Acknowledgments:}
We thank Jonathan Roberts and Ronnie Jansson for their collaboration in the initial phase of this work as well as Jim Matthews for his useful comments.  We also thank our many colleagues in the Pierre Auger Collaboration for their various interactions and input.
A.K. and M.S. acknowledge support from the Department of Energy under grants DE-0009926 and DE-FG02-91-EF0617.
G.F. acknowledges support from the National Science Foundation and NASA under grants NSF-PHY-1212538 and NNX10AC96G; resources supporting this work were provided by the NASA High-End Computing (HEC) Program through the NASA Advanced Supercomputing (NAS) Division at Ames Research Center, consisting of time on the Pleiades supercomputing cluster awarded to G. Farrar.  

\bibliography{references}

\end{document}